\documentclass{PoS}

\usepackage{amsmath,amssymb}

\title{Effects of near-zero Dirac eigenmodes on axial $\mathop{\rm U}(1)$ symmetry at finite temperature}

\ShortTitle{Effects of near-zero Dirac eigenmodes on axial U(1)
}

\author{\speaker{Akio Tomiya}$^1$,
 Guido Cossu$^2$,
 Hidenori Fukaya$^1$,
 Shoji Hashimoto$^{2,3}$,
 Junichi Noaki$^2$\\
        $^1$Department of Physics, Graduate School of Science, Osaka University,
 	Toyonaka, Osaka 560-0043, Japan\\
	$^2$High Energy Accelerator Research Organization (KEK), Ibaraki 305-0801, Japan \\
	$^3$School of High Energy Accelerator Science, The Graduate University for
Advanced Studies (Sokendai), Tsukuba 305-0801, Japan\\
        E-mail: \email{akio@het.phys.sci.osaka-u.ac.jp}}

\newcommand{\Lsym}{ $\mathop{\rm SU}(2)_\text{L}\times \mathop{\rm SU}(2)_\text{R} \times \mathop{\rm U}(1)_V \times \mathop{\rm U}(1)_A$ }
\newcommand{\uva}{$\mathop{\rm U}(1)_\text{V}\times \mathop{\rm U}(1)_\text{A}$ }
\newcommand{\ua}{$\mathop{\rm U}(1)_\text{A}$ }

\newcommand{\su}{$\mathop{\rm SU}(2)_\text{L}\times \mathop{\rm SU}(2)_\text{R}$ }
\newcommand{\mobius }{M\"obius }
\newcommand{\gw}{Ginsparg-Wilson relation}
\newcommand{\tr}{\text{tr}}

\abstract{
We study the axial \ua symmetry of $N_\text{f}=2$ QCD at finite
temperature using the Dirac eigenvalue spectrum.
The gauge configurations are generated employing the \mobius domain-wall
fermion action
on $16^3 \times8$ and $32^3 \times8$  lattices.
The physical spatial size of these lattices is around 2~fm and 4~fm,
respectively,
and the simulated temperature is around 200 MeV,
which is slightly above the critical temperature of the chiral phase
transition.
Although the \mobius domain-wall Dirac operator is expected to have a
good chiral symmetry
and our data  actually show small values of the residual mass,
we observe significant violation of the \gw ~for the low-lying eigenmodes of the \mobius domain-wall Dirac operator.
Using the reweighting technique, we compute the overlap-Dirac operator spectrum
on the same set of configurations and find a significant difference 
of the spectrum between the two Dirac operators
for the low-lying eigenvalues.
The overlap-Dirac spectrum shows a gap from zero,
which is insensitive to the spacial volume.
}

\FullConference{The 32nd International Symposium on Lattice Field Theory,\\
		23-28 June, 2014\\
		Columbia University New York, NY}

\begin{document}
\section{Introduction}
Two-flavor QCD Lagrangian in the vanishing quark mass limit
has  \Lsym symmetries.
Among them, \ua is special because it is violated by the quantization of the theory, {\it i.e.}
the chiral anomaly.
Yet, whether and how anomaly affect the particle spectra are difficult questions.
At zero temperature, it appears as the heavy $\eta'$ mass compared to pions,
while at high temperature it is under active research in the last couple of years.

In this work, we investigate 
the spectral density $\rho(\lambda)$ of the Dirac operator eigenvalue $\lambda$.
It is related to the \ua symmetry through the relation,
\begin{align}
\label{eq:U1sus}
\chi_\pi-\chi_\delta
= \lim_{m\to 0}
\int_0^\infty \!\!\!\!\! d\lambda \; \rho(\lambda)\frac{4m^2}{(m^2+\lambda^2)^2},
\end{align}
where $\chi_\pi$ and $\chi_\delta$ are the susceptibilities of 
the isotriplet pseudoscalar and scalar operators respectively.
When $\chi_\pi-\chi_\delta=0$, the \ua breaking is invisible in the correlators of these channels.
It was argued that if there is a gap in the Dirac spectrum,
{\it i.e.} $\rho(\lambda)=0$ for $\lambda<\lambda_{\rm gap}$
with a finite $\lambda_{\rm gap}$, 
$\chi_\pi-\chi_\delta$ vanishes~\cite{Cohen:1996ng}.
It was further shown that if the \su symmetry is fully restored above the critical temperature,
the Dirac spectrum starts with at least cubic powers of 
$\lambda$ and $\chi_\pi-\chi_\delta$ vanishes under this slightly relaxed assumption~\cite{Aoki:2012yj} . 

The Dirac spectrum can be investigated by numerical simulations of lattice QCD.
The JLQCD collaboration \cite{Cossu:2013uua} and TWQCD
collaboration \cite{Chiu:2013wwa}
reported that \ua symmetry is restored above the critical temperature
using the overlap and the optimal domain-wall
fermions, respectively.
%
On the other hand, LLNL/RBC collaboration \cite{Bazavov:2012qja, Buchoff:2013nra}
and Ohno {\it et. al} \cite{Ohno:2012br} obtained results that suggest the opposite conclusion
using the domain-wall or staggered fermions.
The former two groups employ the fermion action having better chirality,
while the latter two groups performed the simulations on larger volumes.
It is also noted that in \cite{Cossu:2013uua}
the global topological charge was fixed to zero.

In this work, we investigate the systematic effects
which may result in the difference among the previous works,
especially 
between the overlap and domain-wall type fermions. 
There are three possible causes.
The first is the finite volume effect. 
There is always a gap in $\rho(\lambda)$ in 
the finite volume even below $T_c$.
It is therefore important to carefully check the volume
scaling of the gap if it exists.
The second is the accuracy of the chiral symmetry.
As \cite{Aoki:2012yj} suggested,
the full \su symmetry plays a key role to suppress the
\ua breaking effect in the correlators.
The third is the effect of fixing topology.

We perform QCD simulations at around $T=$ 200 MeV (>$T_c$) employing
the  \mobius domain-wall fermion action,
which allows us to simulate QCD on larger volumes than that of the
overlap fermion. The topological charge can change in this formulation. 
We use the code platform IroIro++~\cite{Cossu:2013ola}.
By the \mobius implementation of the domain-wall
Dirac operator, we expect that
the \su symmetry is kept to a good precision.
We also study the effect of small violation of their symmetry 
 by reweighting the \mobius domain-wall Dirac determinant to that of the
overlap Dirac operator.
This reweighting, if realizes, corresponds to the dynamical overlap fermion
simulation without fixing topology.

As we will see below, we found a significant difference
between the \mobius domain-wall and the (reweighted) overlap-Dirac operator spectra.
By checking the chirality of each eigenmode, 
it turned out that the low-modes of the \mobius domain-wall Dirac operator 
violate the Ginsparg-Wilson relation, 
quite significantly 
even when their contribution to the residual mass is small.
Such violation of the Ginsparg-Wilson relation
in the low mode region may have a significant impact in the study of \su and \ua symmetry restoration/breaking.

\section{Lattice setup}

\subsection{Simulation with dynamical \mobius domain-wall quarks}

We employ the \mobius domain-wall fermion action
\cite{Brower:2004xi,Brower:2012vk}
for the quarks.
Its determinant is equivalent (except for overall constants)
to that of a four-dimensional effective Dirac operator
\begin{align}
D^{4D}_\text{DW}(m)=\frac{1+ma}{2}+\frac{1-ma}{2}\gamma_5\text{sgn}_\text{rat}(H_M)
\label{eq:def_dw_eff},&\hspace{5mm}
\text{sgn}_\text{rat}(H_M)=\frac{1-(
T(H_M))^{L_s}}{1+( T(H_M))^{L_s}},
\\
T(H_M)=\frac{1- H_M}{1+ H_M},& \hspace{5mm}
H_M=\gamma_5\frac{2aD_W}{2+aD_W},
\end{align}
where $D_W$ is the Wilson Dirac operator with a negative cut-off scale mass $-1/a$.
We introduce three steps of the stout smearing for the gauge links.
The residual mass, calculated as
\begin{equation}
m_{\rm res}={\frac{\langle\tr G^\dagger\Delta_L G \rangle}{\langle\tr G^\dagger G \rangle}},
\hspace{2mm}
\Delta_L=\frac{1}{2}\gamma_5(\gamma_5 D^{4D}_\text{DW}+D^{4D}_\text{DW}\gamma_5-2 a D^{4D}_\text{DW} \gamma_5 D^{4D}_\text{DW}),
\end{equation}
with $G$ the contact-term-subtracted quark propagator, 
is roughly 5-10 times smaller than
that of the conventional domain-wall Dirac operator for a fixed $L_s$, the size of fifth direction.

For the gauge part, we employ the Symanzik gauge action
with $\beta=4.07$ and $4.10$.
From the measurement of the Wilson flow
at zero temperature 
the lattice spacing is estimated to be 
{$0.135$ fm and $0.125$ fm}, respectively.
For each value of $\beta$, we simulate on two volumes
$L^3\times L_t=16^3\times8$ and $32^3\times8$,
at quark masses
$am_\text{ud}=0.01$ ($30$ MeV or $32$ MeV) and 
$am_\text{ud}=0.001$ ($3.0$ or $3.2$ MeV).  
$L_s$ is chosen such that the residual mass is kept at around or smaller than 1 MeV.
From the Polyakov loop and the chiral condensate,
the simulated temperature, $180$ MeV ($\beta=4.07$)
and $200$ MeV ($\beta=4.10$), is estimated to be
slightly above $T_c$.
For each ensemble, we sample {50-200} 
gauge configurations from {100-700} 
trajectories of the hybrid Monte Carlo updates.

\subsection{The overlap/domain-wall reweighting}
In order to understand
the difference between the domain-wall type fermions
and the overlap fermions, 
we perform the reweighting
of the dynamical \mobius domain-wall ensembles
to those with the overlap Dirac operator determinant.

Our choice of the overlap Dirac operator is obtained by
choosing a better approximation for the sign function
in (\ref{eq:def_dw_eff}),
while keeping the same kernel operator $H_M$. 
On the generated configurations,
we compute lowest eigenmodes $|\lambda_i \rangle$ of
the kernel operator $H_M$, and exactly
calculate the sign function for them.
Namely, we use
\begin{align}
D_\text{ov}(0)=\frac{1}{2}
\sum_{\lambda_i<|\lambda_\text{th}|}(1+\gamma_5 \text{sgn}
\lambda_i)|\lambda_i\rangle\langle\lambda_i|
+D_\text{DW}^\text{4D}(0)
(1-
\sum_{\lambda_i<|\lambda_\text{th}|}|\lambda_i\rangle\langle\lambda_i|
),
\end{align}
where $\lambda_i$ is the  $i$-th lowest eigenvalue of $H_M$ below a threshold $\lambda_\text{th}$.
With our choice $a\lambda_\text{th}=0.35$ (for $L=16$) and
 $0.24$ (for $L=32$) the residual mass is negligible, {\it i.e.} $<4\times 10^{-3}$ MeV.

We perform the overlap/\mobius domain-wall reweighting by computing
\begin{align}
\langle\mathcal{O}\rangle_\text{ov}
=
\left\langle\mathcal{O}
\frac
{\det D_\text{ov}^{2}(m_\text{ud})}
{\det D_\text{DW}^{2}(m_\text{ud})}
\frac
{\det D_\text{DW}^{2}(1/2a)}
{\det D_\text{ov}^{2}(1/2a)}
\right\rangle_\text{DW},
\end{align}
where the ratio of the determinants are stochastically estimated
using $\mathcal{O}(10)$ noise samples for each configuration~\cite{Fukaya-reweighting}.
Here, $\langle \cdots \rangle_{\text{DW}}$ denotes the ensemble average
on the dynamical \mobius domain-wall ensembles.
Note that we have added an additional determinant of fermions and ghosts with a cut-off scale mass $(1/2a)$,
which are irrelevant for the low-energy physics but effective
in reducing statistical fluctuation originating from the UV modes.

It turned out that this overlap/\mobius domain-wall reweighting
is effective only on the smaller lattice ($16^3\times 8$).
On the larger volume {$32^3\times 8$}, 
we instead use the low-mode reweighting, {\it i.e.} approximating the determinants 
by
a product of  lowest $\mathcal{O}(10)$ eigenvalues.
This is not a very precise approximation of the determinant but, as discussed later, 
can still be used to study the possible gap in 
the Dirac eigenvalue spectrum.

\section{Preliminary results}

\subsection{Dirac spectrum}

First, by comparing the spectrum of low-lying eigenvalues of
$\gamma_5D_{DW}(m)$ and that of the reweighted $\gamma_{5}D_{ov}(m)$ 
measured on the same configurations,
we examine the effect of the violation of chiral symmetry. 
Using the ensembles on two different lattice volumes,
we can check the volume scaling at the same time.
Since the configurations are generated with the \mobius domain-wall 
quark action, the topology tunneling is active.

\begin{figure}[tbp]
 \begin{minipage}{0.5\hsize}
  \centering
  \includegraphics[width=8cm]{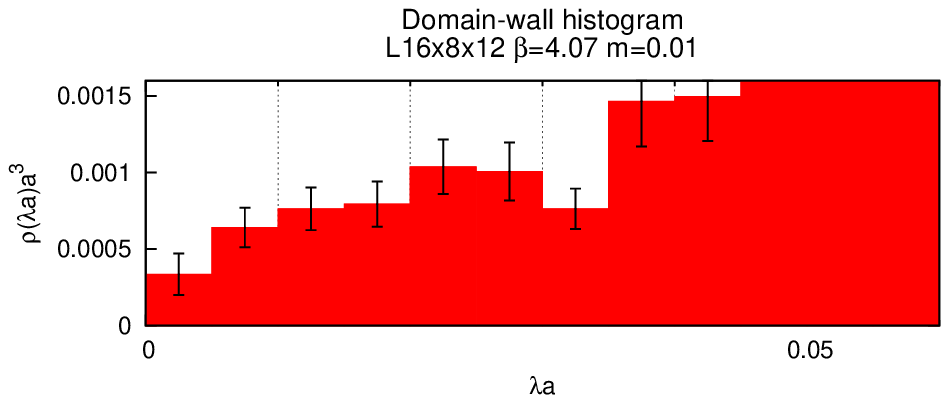}
\label{fig:spectrum2/L16x8b407M100s.eps}
 \end{minipage}
 \begin{minipage}{0.5\hsize}
  \centering
  \includegraphics[width=8cm]{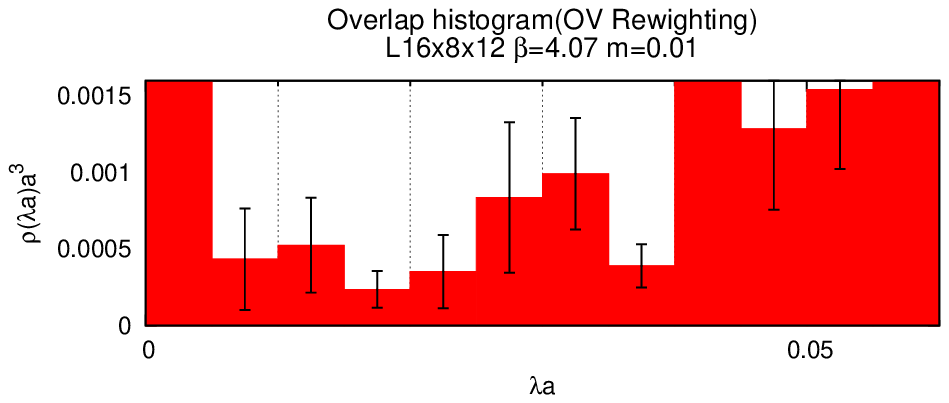}
\label{fig:spectrum2/L16x8b407M100s_ov_rewov.eps}
 \end{minipage}
 \caption{Eigenvalue spectrum of the \mobius domain wall (left panel)
   and reweighted overlap (right) Dirac operators. The data for $am_\text{ud}=0.01$, $T\sim 180$ MeV
   on the  $L^3=16^3\times8$ lattices.
   The peak in the lowest bin in the right panel is $a^3\rho(0)=0.00164\pm 0.00045$.
   \label{DW_heavy_hist}
}
\end{figure}

\begin{figure}[tbp]
 \begin{minipage}{0.5\hsize}
  \centering
  \includegraphics[width=8cm]{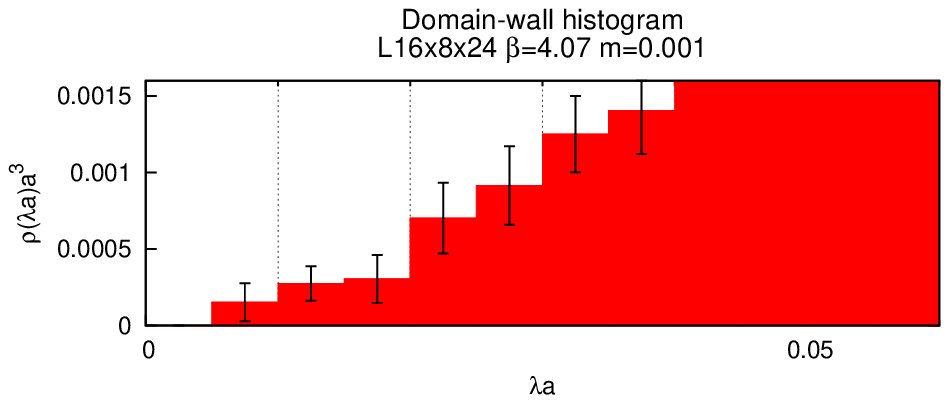}
\label{fig:spectrum2/L16x8b407M100s_m0.001.eps}
 \end{minipage}
 \begin{minipage}{0.5\hsize}
  \centering
  \includegraphics[width=8cm]{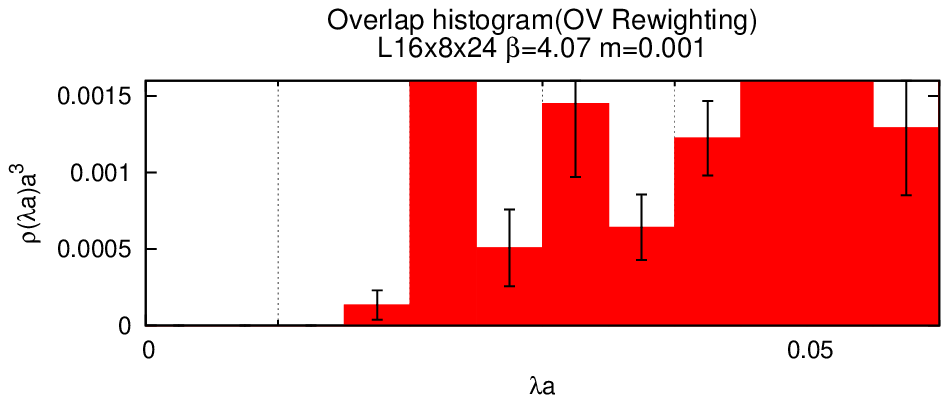}
\label{fig:spectrum2/L16x8b407M100s_m0.001_ov_rewov.eps}
 \end{minipage}
 \\
 \begin{minipage}{0.5\hsize}
  \centering
  \includegraphics[width=8cm]{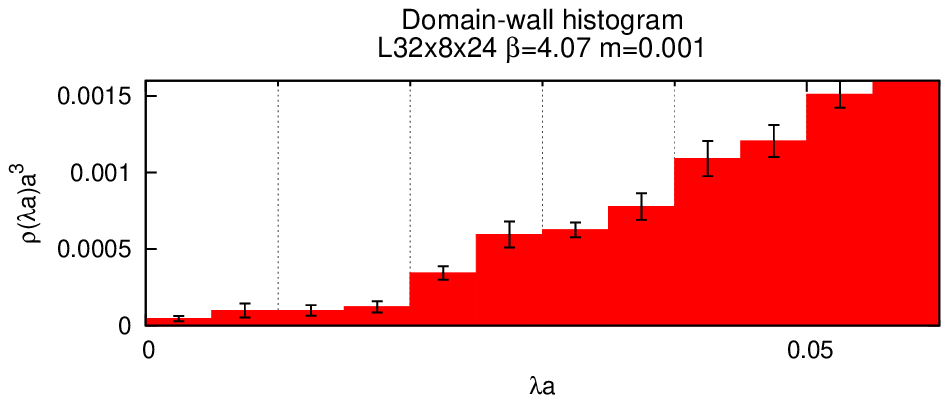}
\label{fig:spectrum2/L32x8b407M100s_m0.001.eps}
 \end{minipage}
  \begin{minipage}{0.5\hsize}
  \centering
  \includegraphics[width=8cm]{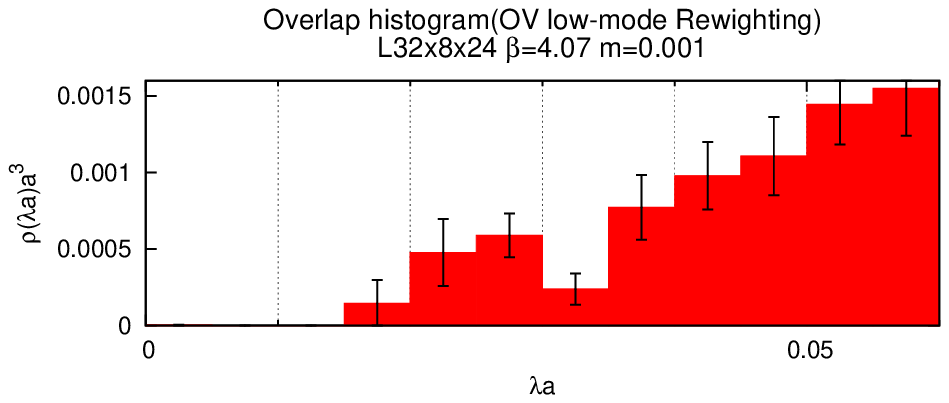}
\label{fig:spectrum2/L32x8b407M100s_m0.001_ov_rewovlow.eps}
 \end{minipage}
 \caption{Eigenvalue spectrum of the \mobius domain wall (left panels)
   and reweighted overlap (right) Dirac operators. The data for $am_\text{ud}=0.001$, $T\sim 180$ MeV
   on the $L^3=16^3\times 8$ (top panels) and  $L^3=32^3\times 8$ (bottom) lattices are presented.\label{DWov_hist}
}
\end{figure}

Fig.~\ref{DW_heavy_hist} and \ref{DWov_hist} show the eigenvalue spectrum $\rho(\lambda)$ calculated on the $T=180$ MeV lattices. 
Here, the $i$--th eigenvalue of massless Dirac operator $\lambda_i$ is obtained by,
\begin{align}
\lambda_i a\equiv \frac{\sqrt{a^2(\lambda_i^m)^2-a^2m_\text{ud}^2}}{\sqrt{1-a^2m_\text{ud}^2}} ,
\end{align}
where $\lambda_i^m$ is  the $i$--th eigenvalue of massive hermitian Dirac operator
$\gamma_5 D^{4D}_\text{DW}(m)$ or $\gamma_5 D_\text{ov}(m)$.
When the quark mass is heavy, $m_\text{ud}\sim30$ MeV,
our data show apparent difference between the \mobius domain-wall and overlap Dirac eigenvalues near $\lambda\sim0$ (Fig. \ref{DW_heavy_hist}).
The left panel shows the data for $\gamma_5D^{4D}_\text{DW}(0)$,
while the right panel is those of (reweighted) $\gamma_5D_\text{ov}(0)$. 
The overlap Dirac spectrum (right panel) has a peak around $\lambda\sim0$, while the \mobius Domain-wall does not.
The peak in the overlap spectrum originates from chiral zero-modes,
which are determined unambiguously thanks to the nearly exact chiral symmetry of the overlap Dirac operator.
Above the peak region, {\it i.e.} $\lambda a\sim$ 0.02, the spectral density for the overlap becomes lower than that of \mobius domain-wall.

On the other hand, for the smaller $m_\text{ud}$ ($\sim$ 3~MeV) we do not find the peak after the reweighting, 
and the near-zero modes around $\lambda a\sim$ 0.01 are washed out as shown in Fig.~2,
where we present the data for $L\sim 2$ fm (top) and $L\sim 4$ fm (bottom).
For the reweighted overlap, a gap $\sim 20$ MeV is found on 
both volumes, while the \mobius domain-wall spectrum shows eigenmodes below $| a\lambda| \approx 0.01$. 
On the large volume,
in particular,
there is an eigenvalue in the lowest bin.
The data at  $T\sim 200$ MeV are qualitatively similar.

The reweighted overlap Dirac spectrum 
shows a gap, which is apparently insensitive to the volume.
We may conclude that the difference from the \mobius domain-wall fermion
is mainly due to the violation of the chiral symmetry,
that we investigate in more detail below.

\subsection{Violation of the Ginsparg-Wilson relation}
We measure the violation of the Ginsparg-Wilson relation on each eigenmode
 of the Hermitian Dirac operator through
\begin{align}
g_i\equiv
\frac{
\psi_i^\dagger\gamma_5[
D\gamma_5 + \gamma_5 D - 2aD \gamma_5 D
]\psi_i}
{\lambda_i^m}
\left[
\frac{(1-am_\text{ud})^2}{2(1+am_\text{ud})}
\right]\label{eq:def_gi},
\end{align}
where $\lambda_i^m$, $\psi_i$ denotes the $i$--th eigenvalue/eigenvector of massive hermitian Dirac operator respectively.
Last factor in (\ref{eq:def_gi}) comes from the normalization of the Dirac operator.
Note that one can obtain the residual mass by an weighted average of $g_i$,
\begin{align}
m_\text{res}
={\frac{\langle\tr G^\dagger\Delta_L G \rangle}{\langle\tr G^\dagger G \rangle}}
=
\left.\sum_i\frac{\lambda^m_i(1+am_\text{ud})}{(1-am_\text{ud})^2(a\lambda^m_i)^2}g_{i}\right/
\sum_i\frac{1}{(a\lambda^m_i)^2}.
\end{align}
\begin{figure}[ptb]
  \centering
 \includegraphics[width=11cm]{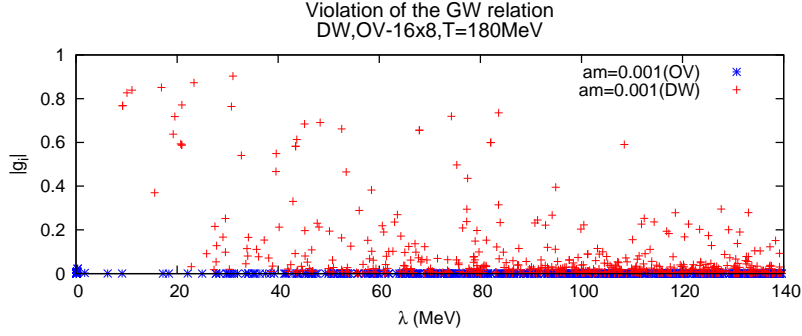}
\caption{Violation of the Ginsparg-Wilson relation $g_i$ for individual eigenmode.
Pluses represent the \mobius domain-wall eigenvectors, while stars show the overlap eigenvectors, which are of course zero.}
\label{fig:gi/gi_smeared_SymDW_sHtTanh_16x8x24_b4.07_M1.00_mud0.001}
\end{figure}
where the sum runs over all eigenvalues.

Figure \ref{fig:gi/gi_smeared_SymDW_sHtTanh_16x8x24_b4.07_M1.00_mud0.001} shows $|g_i|$ for each eigenvalue
on the configuration of $16^3\times8$ and $m_\text{ud}\sim3$ MeV.
For the \mobius domain-wall fermion (crosses),
the low-lying modes violate the chiral symmetry to the order of one, which means that the expectation value of 
$D\gamma_5 +\gamma_5 D -2 a D\gamma_5 D$ is of the same order of $\lambda$.
The violation is of course negligible for the overlap fermion (stars).
This result indicates that the low modes of the \mobius domain-wall Dirac operator
contain substantial lattice artifact. 
Such lattice artifacts may also distort the eigenvalues, 
and explain the difference from the overlap operator.

\subsection{Low mode reweighting}
As mentioned above, the conventional stochastic reweighting
does not work on the larger lattice.
Instead, we introduce an approximation of using only the low-lying eigenvalues.
This corresponds to a certain modification of the fermion action in the ultraviolet regime.
We incorporate all the eigenvalues below $\lambda\sim 100$ MeV. 
Here, we show that this low-mode reweighting can be used to
study the gap in the Dirac spectrum.

\begin{figure}[ptb]
  \centering
  \includegraphics[width=11cm]{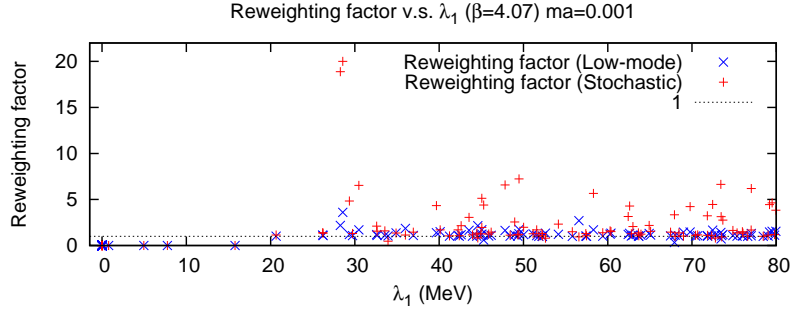}
\caption{
Reweighting factor with the low-mode reweighting (stars) and stochastic reweighting including all modes (pluses).
The horizontal axis is the lowest eigenvalue of the overlap-Dirac operator for that gauge configuration.
Data at $\beta=4.07$ and $am_\text{ud}=0.001$ on the $16^3\times8$ lattice are plotted.
}
\label{fig:reweighting/eigen_rew_HovTanhthre0.35_Beta4.07_m0.001.eps}
\end{figure}

On the smaller lattice, we compare the reweighting and the low-mode reweighting as shown in Fig.~\ref{fig:reweighting/eigen_rew_HovTanhthre0.35_Beta4.07_m0.001.eps}.
Pluses and crosses represent the conventional stochastic reweighting factor and the low-mode reweighting factor, respectively.
Each point represents a gauge configuration on which the reweighting factor is calculated.
As the horizontal axis, we take the first eigenvalue $\lambda_1$.
Below $\lambda_1\sim 20$ MeV, both reweighting factors are consistent and essentially zero.
Configurations having near-zero modes are strongly suppressed in both reweighting techniques,
and we may therefore conclude that the non-existence of the gap in the Dirac spectrum
does not depend on the details of the reweighting technique.

\section{Summary}
We have studied the low-lying eigenvalue spectrum
of the \mobius domain-wall and reweighted overlap Dirac operators
slightly above the critical temperature.
Our preliminary result at the lightest quark mass 
shows a significant difference between them.
The overlap-Dirac eigenvalue spectrum for the lightest quark mass shows a gap, which  is insensitive to the volume, 
while that of the \mobius domain-wall has small but non-zero spectrum near $\lambda=0$.
The large violation of the Ginsparg-Wilson relation on the low-modes of the domain wall operator
may explain the difference.
\vspace{2mm}

Numerical simulations are performed on IBM System Blue Gene Solution at KEK under a support of its Large Scale Simulation Program (No. 13/14-04).
We thank H. Matsufuru for the support on the computing facility and P. Boyle for helping in the optimization of the code for BG/Q. This work is supported in part by the Grand-in-Aid of the Japanese Ministry of Education (No.25800147, No.26400259) 
and SPIRE (Strategic Program for Innovative Research) Field 5.

\end{document}